\documentclass{ws-ijmpe}
\usepackage{epsfig}

\begin{document}

\markboth{A. Baran, Z. \L ojewski and K. Sieja}
         {State dependent $\delta$-pairing and spontaneous fission}
\author{\footnotesize A. BARAN, Z. {\L}OJEWSKI and K. SIEJA}
\address{Institute of Physics, University of M. Curie-Sk\l odowska, \\ 
ul. Radziszewskiego 10, 20-031 Lublin, Poland}
\title{\bf STATE DEPENDENT $\delta$-PAIRING AND SPONTANEOUS FISSION}

\maketitle

\begin{history}
\received{September 22, 2003}
\revised{October 30, 2003}
\end{history}

\begin{abstract}
We examine the fission barriers, mass
parameters and spontaneous fission half lives of fermium isotopes
within the framework of macroscopic-microscopic model with a
$\delta$-pairing interaction.
Four different macroscopic models are applied and studied.
The results are compared to experimental data and to the corresponding 
monopole pairing ones. The half lives obtained in the $\delta$-pairing 
model are comparable with experimental data. The state-dependence
of pairing has an important effect on the calculated fission half-lives.
\end{abstract}

The calculations of fission half lives $T_{sf}$ in 
macroscopic-microscopic methods consist of determining the fission barrier
(collective potential energy) and the mass tensor of the nucleus
undergoing the fission process.
The potential energy $V$ splits into a shell $\delta{}E_{\rm shell}$ and
a pairing correction $\delta{}E_{\rm pair}$ (microscopic part) and the 
smooth average (macroscopic) energy 
such as liquid drop\cite{MS67}, droplet\cite{MS74}, LSD\cite{Pomorski} 
or Yukawa plus Exponential.\cite{Krap79}

The microscopic part of the energy is calculated
using the single-particle Woods-Saxon potential 
with the {\it universal} set of parameters.\cite{Cwiok87}

All the former calculations  of $T_{\rm sf}$ were done assuming
that the pairing matrix elements 
$G_{ab}\equiv \langle\nu\bar\nu|\hat V_{\rm pair}|\mu\bar\mu\rangle$ 
are constant. 
Such an approximation leads to the averaging of the superconductive properties 
of nuclei and weakly reflects the structure of nucleon pairs.
In the following we have used the state dependent pairing model
where the pairing matrix elements $G_{ab}$
are state dependent and the interaction 
$\hat V_{\rm pair}$ is of the form:\cite{Krieger}
\begin{equation}
\hat V_{\rm pair}=-V_0\frac{1-\sigma_1\cdot\sigma_2}{4}\delta(\vec r_{12})\,.
\end{equation}
Here $V_0$ is a constant determined from experimental nuclear masses.\cite{sieja,Wapstra}
\begin{figure}[htb]
\centerline{\epsfig{file=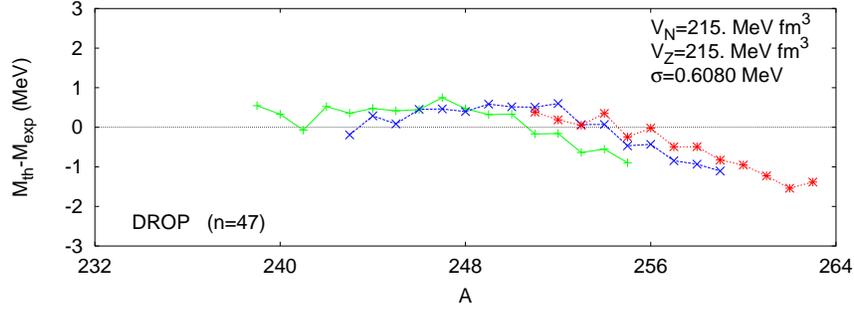,height=4.5cm,}}
\caption{Mass deviations for 47 isotopes of Z=99, 100, 101
in the case of the state-dependent $\delta$-pairing and the liquid drop model.\label{fig-mass}}
\end{figure}

This form of the interaction better reproduces the properties 
of the nucleons  coupled in pairs in degenerate time reversed states.
The $\delta$-pairing 
strength constants $V_0$ for protons ($p$) and neutrons ($n$) are 
$V_0^p = V_0^n = 215\, {\rm MeV\, fm}^3$.
The masses\cite{Wapstra}
in the vicinity of the fermium nuclei are reproduced in 
this case with an accuracy given by the standard deviation error $\sigma=0.608$MeV 
(see Fig. \ref{fig-mass}).
The residual pairing interaction is treated in the BCS approximation. 

Spontaneous fission of a nucleus is described as a tunnelling of
the collective potential barrier described in WKB approximation. In order to
minimize the action which enters the fission probability
we have used the dy\-na\-mic pro\-gra\-mming method.\cite{Baran3,Staszczak} 

The ma\-cro\-sco\-pic-micro\-sco\-pic method is not analytical. 
Therefore, it is necessary to calculate the potential energy and all the components 
of the tensor of inertia $B_{kl}({\rm def})$ on a~mesh in the 
mul\-ti-di\-men\-sio\-nal space spanned by a set of collective degrees of freedom 
consisting of three nuclear shape parameters ($\beta_2$, $\beta_4$, $\beta_6$).
The collective mass $B_{kl}({\rm def})$ is calculated in the adiabatic cranking 
model.

\begin{figure}[htb]
\centerline{\epsfig{file=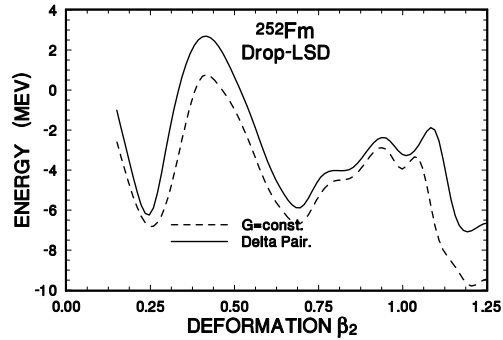,height=4.5cm,angle=0}}
\caption{The fission barriers for $^{252}$Fm in case of both
$G={\rm const}$ (dashed curve) and $\delta$-pairing (solid curve) models.
\label{fig-barriers}}
\end{figure}

An example of fission barriers of $^{252}$Fm evaluated 
in both monopole and $\delta$- pairing models is shown in Fig.~\ref{fig-barriers}. 
The monopole pairing barrier ($G={\rm const}$) is denoted by the dashed 
line and the $\delta$-pairing one is represented by the solid curve. 
One observes that the monopole pairing  model gives a fission barrier 
similar to that of the $\delta$ pairing force model.
However, the barrier height is smaller for $G={\rm const}$ pairing.
\begin{figure}[htb]
\centerline{\epsfig{file=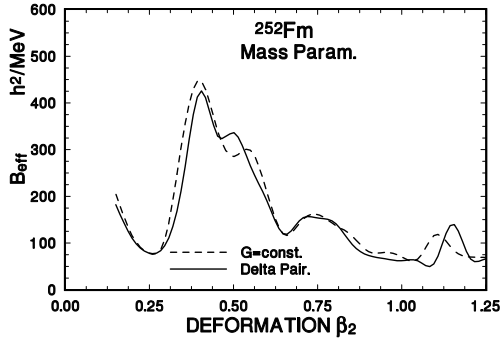,height=4.5cm,angle=0}}
\caption{The effective mass parameter for $^{252}$Fm for two different pairing
models, $\delta$ (solid) and $G={\rm const}$ (dashed curve) models of pairing.
\label{fig-figmas}
}
\end{figure}

Figure \ref{fig-figmas} shows the effective mass parameters taken along the path to 
fission. One can see the differences of both monopole pairing 
(dashed curve) and $\delta$-pairing (solid curve) effective inertia.

\begin{figure}[htb]
\begin{center}
\psfig{file=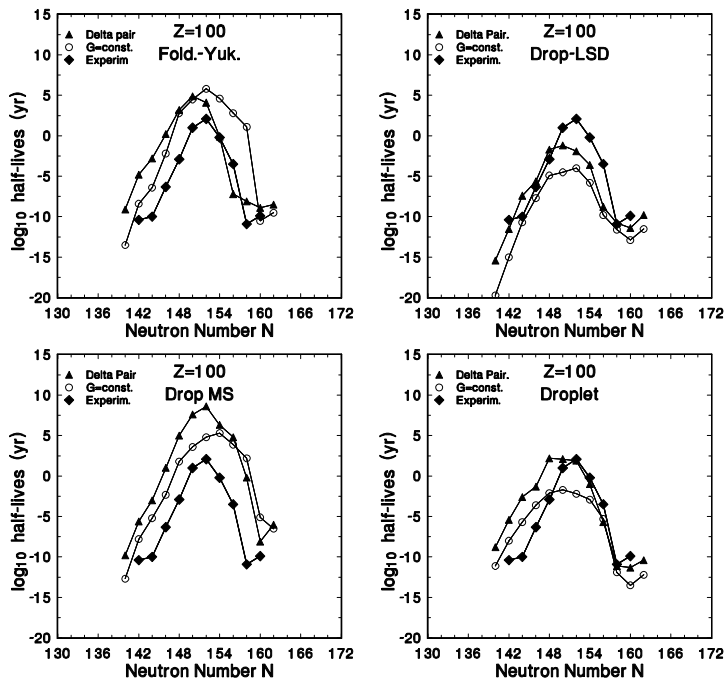,height=10cm,angle=-90}
\end{center}
\caption{Spontaneous fission half-lives ($T_{\rm sf}$) of fermium isotopes 
in four models of the macroscopic energy and two models of pairing interactions. 
The results  of the $\delta$ pairing model are denoted by plain triangles and the results
of the monopole pairing are represented by open circles. 
Diamonds correspond to the experimental data.\label{fig-figtsf}}
\end{figure}

As was shown the fission barriers strongly depend on the 
macroscopic energy model.\cite{loe} To examine this fact in the case of the $\delta$-pairing 
model we have studied  four most frequently used macroscopic models.  
The choice of the model strongly influences the spontaneous fission 
half-lives $T_{\rm sf}$. 
This is illustrated  in Fig. \ref{fig-figtsf} where the results of calculations 
 of the spontaneous fission half-lives for the fermium isotopes  are presented.
The Figure shows the results for the Yukawa plus exponential macroscopic energy, 
Lublin-Strasbourg  Drop (LSD), the Mayers-Swiatecki drop and for the droplet model.
Triangles correspond to the $\delta$-pairing model and the open circles
to the monopole pairing one. The experimental data are denoted by plain diamonds. 
It is seen that the spontaneous fission half-lives considerably depend on the 
pairing model. 
The $\delta$-pairing interaction seems to work better with most macroscopic models 
as compared to the monopole pairing model.

In our comparative study we have neglected the lowest odd-multipolarities 
usually represented by $\beta_3$ and $\beta_5$ deformations which, as was
shown elsewhere,\cite{loe1} considerably decrease the fission barriers especially 
at large quadrupole deformations. 

From our investigations one can draw the following conclusions:
\begin{romanlist}
\item The fission barriers of the $\delta$-pairing force model 
are similar to that of the clasical pairing, however slightly higher. 

\item
The state dependent $\delta$-type force influences  significantly  
the spontaneous fission half lives. At the same time the isotopic systematics
of $T_{\rm sf}$ does not change.

\item The $T_{\rm sf}$ values  calculated in different macroscopic models are
considerably different and depend on the pairing model used. 

\item The investigations of the pairing interaction
and the macroscopic energy models should be continued,
in order to obtain the most
appropriate fission barriers and spontaneous fission half lives. 
\end{romanlist}

\end{document}